\documentclass{article}
\usepackage{amssymb,amsmath,delarray,epsfig}
\begin{document}

\begin{center}
{\LARGE \bf Computing fractal dimension in supertransient systems directly, fast and reliable}
\vskip.25in
{{\large Romulus Breban$^*$ and Helena E. Nusse$^\dagger$}\vskip.1in
{\it $^*$Semel Institute for Neuroscience and Human Behavior, University of California, Los Angeles, California 90024, USA;\\
$^\dagger$University of Groningen, Department of Econometrics, P.O.~Box 800, NL-9700 AV, Groningen, The Netherlands.}}
\end{center}
\vskip.35in

\begin{abstract}
Chaotic transients occur in many experiments including those in fluids, in simulations of the plane Couette flow, and in coupled map lattices and they are a common phenomena in dynamical systems. Superlong chaotic transients are caused by the presence of chaotic saddles whose stable sets have fractal dimensions that are close to phase-space dimension. For many physical systems chaotic saddles have a big impact on laboratory measurements, and it is important to compute the dimension of such stable sets including fractal basin boundaries through a direct method. In this work, we present a new method to compute the dimension of stable sets of chaotic saddles directly, fast, and reliable.
\end{abstract}

Chaotic transients are common phenomena in dynamical systems. They were observed in the Lorenz system \cite{KY_YY1,KY_YY2}, in fluid experiments \cite{AW_BD_DM1,AW_BD_DM2,AW_BD_DM3}, in many low-dimensional dynamical systems \cite{GOY1_GOY2_GOY31,GOY1_GOY2_GOY32,GOY1_GOY2_GOY33,KG}, in spatiotemporal chaotic dynamical systems \cite{CK,LW}, simulations of the plane Couette flow (a shear flow between two parallel walls) \cite{SE_EM_EMS1, SE_EM_EMS2}, and in coupled map lattices \cite{K}. Generally, these transients are caused by the presence of chaotic saddles \cite{NY1,NY2,NY3,Twl} (i.e., strange repellers). When there is a chaotic saddle in the phase space, trajectories originating from random initial conditions in a neighborhood of the stable set of the chaotic saddle (such as in the neighborhood of a fractal basin boundary (BB)), usually wander in the vicinity of the chaotic saddle for a finite amount of time before escaping the neighborhood of the chaotic saddle and settling into a final attractor. Chaotic saddles can strongly effect observed experiments and a ``turbulent'' state in experiments could be supported by a chaotic saddle rather than an attractor. If there are two or more attractors in the system, then chaotic saddles may be buried in the boundaries of basins of attraction leading to fractal BBs. Supertransients are superlong (chaotic) transients and occur commonly in spatiotemporal chaotic dynamical systems \cite{CK,K,LW}. In such a case, trajectories starting from random initial conditions in a neighborhood of the stable set of a chaotic saddle wander chaotically for a long time before settling into a final attractor. It was observed in numerical experiments that spatially extended systems exhibit supertransients so that the observation of the system's asymptotic attractor is practically impossible \cite{CK,K}. Reference \cite{LW} investigates the geometric properties of the chaotic saddles which are responsible for the supertransients in spatiotemporal chaotic systems, and reports that the stable set of the chaotic saddle possesses a fractal dimension which is close to the phase-space dimension. There are a few tools for computing the dimension of 
(one-dimensionally unstable) stable sets of chaotic saddles \cite{KG,HGO,NY1,NY2,NY3,HOY,LW}. These methods involve the computation of Lyapunov exponents. A frequently used method is the combination of the Sprinkle method \cite{KG} and the method for computing the maximal Lyapunov exponent of a typical Saddle Straddle Trajectory \cite{NY1,NY2,NY3}. However, computing the Lyapunov exponents of a Saddle Straddle Trajectory in a supertransient system is extremely time consuming. 

\begin{figure}[t]
\begin{center}
\mbox{\epsfig{file=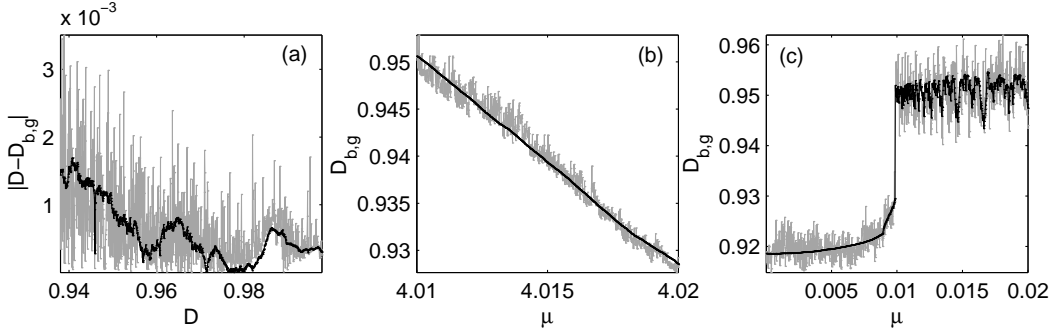,width=14cm}}
\caption{(a) $|D_{g}-D|$ versus $D$ (black) and $|D_{b}-D|$ versus $D$ (gray), where $D_g$ ($D_b$) is the numericaly calculated grid (box) dimension, and $D$ is the dimension calculated analytically for our first numerical example. We used $\bar n=5$, $\bar k=17$, and $r=3$; (b) $D_{g}$ versus $\mu$ (black) and $D_{b}$  versus $\mu$ (gray), where $D_g$ ($D_b$) is the numericaly calculated grid (box) dimension, and $\mu$ is the parameter of our second numerical example. We used $\bar n=5$, $\bar k=17$, and $r=3$. The refinements in (a, b) are tested with the Straddle test. Our numerics clearly illustrate a precision gain up to an order of magnitude when the dimension is calculated by our method as compared to when the dimension is calculated by the box counting algorithm.  Both calculations use the same grid of points so they are equally fast. Indirect methods involving the calculation of Lyapunov exponents are much slower; (c) Similar to (b) for the last numerical example except that $\bar n=5$, $\bar k=19$, and $r=1$. The refinements are tested with the BB straddle test.}
\label{f:f2}\end{center}
\end{figure}

The purpose of this Letter is to present a simple, general method (involving no Lyapunov exponents) for computing the dimension of the stable set of a chaotic saddle that is responsible for the supertransients and the computation is direct, fast and reliable. We refer to this algorithm as the ``Straddle Grid Dimension Algorithm'' (SGDA); it can be applied to flows (differential equations) or maps. Our main result is that the SGDA can be applied to supertransient systems having one-dimensionally unstable stable sets such as basin 
boundaries in $d$-dimensional space ($d\geq 1$). To illustrate the effectiveness of our method for computing dimension see Fig.~\ref{f:f2}.  As indicated by our numerics, the SGDA is superior to applying the box-counting algorithm (BCA) by utilizing straddle pairs for the end points of the boxes. 
     
We present the SGDA for maps but only trivial modifications are necessary for flows \footnote{To apply the method to flows, consider the map generated by the numerical solver which takes the system from a point in the phase space at time $t_n$ to a point at time $t_{n+1} = t_n + \Delta_{t_n}$, and apply the method.}. Let $f$ be a continuous map from the $d$-dimensional phase space $M$ into itself such that $f$ is a $C^3$-diffeomorphism when $d \geq 2$, or a $C^2$-map when $d=1$. Let $R$ be a ``transient region'' in $M$ (that is, a region in the phase space that contains no attractor; of course, a transient region may contain points that belong to an attractor). Under some  mild assumptions, almost all orbits with initial conditions in the transient region leave this region in a finite number of iterates under $f$ and approach  one of the attractors of $f$. The {\it escape time} $T(x)$ of $x$ from $R$ is the minimum integer $n \geq 0$ for which the $n$th iterate of $x$ is not in $R$; write $T(x)=\infty$ if all forward iterates of $x$ are in $R$. The escape time of points in the transient region is at least 1, so for $x\in R$, one has $T(x)\geq 1$. Let $C$ be the largest invariant set of $f$ in $R$, that is, $f(C) = C$, so $T(x)=\infty$ for $x \in C$. Assume that $C$ is nonempty and $T(x)$ is finite for almost every point $x \in R$. The maximal invariant $C$ can contain complicated chaotic sets but it can be just a periodic orbit or a fixed point. The stable set of $C$, $S_c$, is the collection of all points $x \in R$ such that $T(x)=\infty$. If the escape time $T(x)$ of a point $x$ is high, then the initial condition $x$ is close to $S_c$. 

Before we present our method, we first discuss the usual indirect method for computing the dimension of stable sets. Assume that $C$ is a chaotic  saddle in the transient region $R$. To see what happens  for  a  uniform  distribution of initial points, assume that at $t = 0$, one selects $N(0)$ initial points (for $N(0)$ large) from  a uniform distribution on $R$. Evolve these  $N(0)$  initial  conditions  under  the dynamics. For $t\geq 0$, let $N(t)$ denote the number of trajectories that are still in the region $R$ at time $t$. Typically, $N(t)$ decays exponentially as time $t$ increases \cite{PY}. Let $A_t$  be the set of points $x\in R$  with  escape time $T(x)\geq t$. Write $m(A_t)$ for the Lebesgue measure of $A_t$. Typically, $m(A_t)\rightarrow 0$ exponentially fast as $t \rightarrow \infty$; that is, for large $t$, $m(A_t)\approx K\exp(-t/\tau)$, for some positive constants $K$ and $\tau$;  ($\tau$ is called the average escape time \footnote{An average escape time $t$ of the chaotic transients can be  defined  by $1/\tau = \lim_{t\to\infty}\lim_{N(0)\to\infty}\frac{1}{t}\ln\frac{N(0)}{N(t)}$; \cite{HGO,HOY,PY}.}). For two-dimensional systems , the dimension $d_u$  of the collection of points of intersection of a line segment with the stable set $S_c$  can be computed indirectly from the relation $\tau \approx 1/(1-d_u)\lambda_1$ \cite{KG,HGO}, where $\lambda_1$ is the maximum Lyapunov exponent for typical trajectories on the chaotic saddle $C$. By using the Sprinkle method \footnote{The Sprinkle method starts with $N$ initial points $x$ that are uniformly distributed in the transient region. Given two integers $N_i$ and $N_f$, for each initial point $x$ with  $T(x)>(N_i+N_f)$, the first $N_i$ iterates of $x$ are discarded and so are the last $N_f$; the remaining iterates are plotted. The accuracy and the speed of the Sprinkle method for finding points close to the chaotic saddle are limited \cite{KG}.}, one computes $\tau$, and by using the PIM-triple method \footnote{Unlike the Sprinkle method, the PIM-triple method quickly generates trajectories which are very close to the stable set of the chaotic saddle, and Lyapunov exponents can be computed fast and accurately for non-supertransient systems \cite{NY1,NY2,NY3}.}, one computes $\lambda_1$. Thus, it is important to have a general method that computes such dimensions directly, fast and reliable. 

We now introduce some concepts needed for describing our new method.  For points $a,b\in M$, and $\epsilon>0$, we call $\{a,b\}$ an $\epsilon$-{\it pair} of points if the distance $||a-b|| = \epsilon$, and a {\it straddle pair} if the line segment $[a,b]$ contains points of $S_c$. From the computational point of view, to decide if a pair $\{a,b\}$ is a straddle pair, one performs a finite (and usually small) number of computations studying a refinement of the pair $\{a,b\}$. A $\delta$-{\it refinement} of the pair $\{a,b\}$ is a finite set of points $X=\{X_i\}_{i=0}^s$ in the segment $[a,b]$ with $X_0=a$ and $X_s=b$ and the distance between two consequtive points $X_i$ and $X_{i+1}$ is $\delta||a-b||$. In our method for computing the dimension of stable sets of a chaotic saddle, we make use of a test (based on the PIM-triple 
concept \cite{NY1,NY2,NY3}), for verifying whether a pair $\{a,b\}$ is a straddle pair. A triple $\{X_i,X_j,X_k\}$ of points in a typical line segment in $M$ intersecting $S_c$ is called a {\it PIM triple} if (a) $X_j$ is between $X_i$ and $X_k$, and (b) the escape time of $X_j$ is bigger than both the escape times of $X_j$ and $X_k$, that is, $T(X_j)>T(X_i)$ and $T(X_j)>T(X_k)$. The ``Straddle test'' below is a basic tool for the SGDA.

{\it Straddle test.} Let $I$ denote a typical line segment in $M$ intersecting $S_c$ and $\epsilon>0$ be given. Let $\{a,b\}$ be an $\epsilon$-pair in $I$ and $X=\{X_i\}_{i=0}^s$ be a refinement of the pair $\{a,b\}$ in $I$. If $X=\{X_i\}_{i=0}^s$ includes a PIM triple then $\{a,b\}$ is a straddle pair \footnote{If the stable set in $R$ is a BB (that is, a point $x \in M$ is on the BB if every neighborhood of $x$ contains points of at least two different basins), then instead of the above straddle test, we may choose to apply the following ``BB straddle test''. 
{\it BB straddle test.}  Let $I$ denote a typical line segment in $M$ intersecting $S_c$ and $\epsilon>0$ be given. Let $\{a,b\}$ be an $\epsilon$-pair in $I$ and $X=\{X_i\}_{i=0}^s$ be a refinement of the pair $\{a,b\}$ in $I$. If there exist $0\leq i,j\leq s$ such that $X_i$ and $X_j$ belong to different basins, then the pair $\{a,b\}$ is a straddle pair.}.

Note that the above procedure tests only finite sets of points; even if the Straddle test does computationally diagnose a pair $\{a,b\}$ as not a straddle pair for a given refinement, this does not imply that the pair $\{a,b\}$ is not a straddle pair. Quite possibly, a different choice of the refinement of $\{a,b\}$ may reveal that $\{a,b\}$ is a straddle pair. Thus, computationally, one always underestimates the number of straddle pairs. The BCA has specific procedures for selecting line intervals of size $\epsilon$ in $I$. Then, the ends of the intervals are chosen as $\epsilon$-pairs, and each $\epsilon$-pair may be further refined. The BCA evaluates the fraction of $\epsilon$-pairs that are straddle pairs according to the Straddle test, where $\epsilon$ decreases on a logarithmic scale. The fraction of straddle $\epsilon$-pairs detected computationally is used as an estimate of the fraction of line intervals of size $\epsilon$ that straddle the fractal set. Thus, the fact that every test based on a finite number of computations does not accurately detect straddle pairs is a major computational difficulty for the BCA. 
\begin{figure}[t]
\begin{center}
\mbox{\epsfig{file=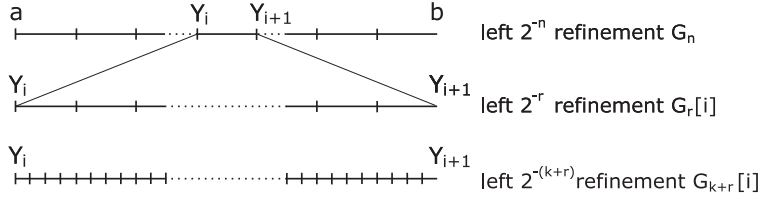,width=10cm}}
\caption{The left refinements (grids) used in the selection procedure.  The points of the refinements are marked by short vertical segments.}
\label{f:f1}\end{center}\vspace{-8mm}
\end{figure}
We now present the SGDA which allows us to compute the dimension of the stable set of a chaotic saddle. Given a grid of points in $I$, we use the grid points to construct $\delta$-refined $\epsilon$-pairs that we test with the Straddle test. We denote the fraction of the selected pairs that are straddle pairs by $\psi_\epsilon$. Similar to the scaling of fractal and uncertainty dimensions \cite{McD,NY2_}, using an approach similar to \cite{NY2_} guaranteeing that these types of dimensions of certain basin boundaries coincide, we argue that $\psi_\epsilon\sim\epsilon^{1-D_g}$ for small $\epsilon$ \footnote{The probability that an interval of size $\epsilon$ contains a straddle pair of any size less than $\epsilon$ must equal the probability that the interval straddles the fractal set. Using Lebesgue measure to {\it count} the number of $\delta$-refinements, we have $\phi_\epsilon=\left[\int_0^\epsilon(\epsilon-\delta)\psi_\delta d\delta\right]/\left[\int_0^\epsilon (\epsilon-\delta)d\delta\right]=(2/\epsilon^2)\int_0^\epsilon (\epsilon-\delta)\psi_\delta d\delta.$
Taking two derivatives of $\epsilon^2\phi_\epsilon$ with respect to $\epsilon$ we obtain
$\psi_\epsilon=\phi_\epsilon+\epsilon\partial_\epsilon\phi_\epsilon+\epsilon^2\partial^2_\epsilon\phi_\epsilon/2$, and thus, since $\phi_\epsilon\sim \epsilon^{1-D}$ so does $\psi_\epsilon$.}, where $D_g$ is by definition the {\it straddle grid dimension}.  An important ingredient for our algorithm is the selection procedure for $\epsilon$-pairs and $\delta$-refinements. In the ``Selection procedure'' below, we consider {\it left} $\delta$-refinements of $\{a,b\}$; that is, a $\delta$-refinement of $\{a,b\}$ in which the right point $b$ is excluded. Let $a$ and $b$ be the end points of $I$. For fixed nonnegative integers $n$, $k$, and $r$, start with a grid $G_{n+k+r}$ which is a left $2^{-(n+k+r)}$-refinement of $\{a,b\}$. Before we state the selection procedure, we specify subgrids to be used. $G_n$ is the left $2^{-n}$-refinement of $\{a,b\}$ included in $G_{n+k+r}$. Denote the grid points of $G_n$ by $\{Y_i\}_{i=1}^{2^n}$, where $Y_1 = a$ and $Y_i<Y_{i+1}$ for all $i$.  For all $i$, write $G_{k+r}[i]$ for the left $2^{-(k+r)}$-refinement of $\{Y_i,Y_{i+1}\}$,  $G_r[i]$ for the left $2^{-r}$-refinement of $\{Y_i,Y_{i+1}\}$ (see Fig.~\ref{f:f1}), and $S^j(G_r[i])$ for the resulting subgrid of $G_{k+r}[i]$ when $G_r[i]$ is shifted $j$ grid steps to the right. The union of the $S^j(G_r[i])$'s ($0\leq j< 2^k$) is $G_{k+r}[i]$. 

{\it Selection procedure for $\epsilon$-pairs and $\delta$-refinements.}

1. Start with the left $2^{-n}$-refinement of $\{a,b\}$ which is the grid $G_n$. Initialize $i=1$. (Of course, here $\epsilon = 2^{-n}$.)
 
2. Start with the $2^{-n}$-pair $\{Y_i,Y_{i+1}\}$, the left $2^{-r}$-refinement of $\{Y_i,Y_{i+1}\}$ (grid $G_r[i]$), and the left $2^{-(k+r)}$-refinement of $\{Y_i,Y_{i+1}\}$ (grid $G_{k+r}[i]$); see Fig.~\ref{f:f1}. 

3. Select the subgrids $S^j(G_r[i])$ in the grid $G_{k+r}[i]$ for $0\leq j<2^k$.

4. Increase $i$ by one. Repeat steps 2., 3., and 4.~for all $1<i\leq 2^n$.

When shifting the grid, the resulting grid $S^j(G_r[i])$ is not a refinement of $\{Y_i,Y_{i+1}\}$ but it is a left $2^{-r}$-refinement of $\{S^j(Y_i),S^j(Y_{i+1})\}$. The above procedure selects a total of $2^{n+k}$ refinements of $2^r$ points each. Every point of the grid $G_{n+k+r}$ is considered exactly once in the selected refinements.  

{\bf Straddle Grid Dimension Algorithm.} \footnote{A fractal BB is a one-dimensionally unstable stable set of a chaotic saddle. For computing the dimension of fractal BB, instead of the box-counting method, one may apply the Uncertainty Dimension Algorithm (UDA) \cite{McD}. If the stable set is a BB, one may apply the BB straddle test instead of the Straddle test. If $I$ intersects a BB, then step (ii) in the SGDA may be replaced by ``For each point $x \in G_{n+k+r}$ except the right point of $I$ determine the basin the point $x$ is in.'' and the first part of step (iii) may be replaced by ``Apply the selection procedure for the integer values $n$, $k$, and $r$ and then the BB straddle test.''  When the stable set is a fractal BB, SGDA is superior to UDA.} 

(i) For three integers $\bar n$, $\bar k$, and $r$, consider a grid $G_{\bar n+\bar k+r}$ which is a left $2^{-(\bar n+\bar k+r)}$-refinement of $\{a,b\}$ where $a$ and $b$ are the end points of $I$. Initialize $k=\bar k$ and $n=\bar n$; 

(ii) For each point $x\in G_{\bar n+\bar k+r}$ compute the escape time $T(x)$ from the region $R$;

(iii) Apply the selection procedure for $n$, $k$, and $r$. Using the Straddle test, count the number of the selected refinements whose pairs straddle $S_c$. Denote the ratio of this number of selected refinements to the total number of refinements by $\psi_n$ ($\equiv\psi_\epsilon$ where $\epsilon=2^{-n}$); 

(iv) Increase $n$ by $1$, decrease $k$ by $1$. If $k<0$ proceed to (v), else proceed to (iii); 

(v) Plot $\log(\psi_n)$ versus $n \log 2$. The slope of this graph is $1-D_g$, where $D_g$ is the straddle grid dimension on $I$. 

Denote the fraction of line intervals of size $2^{-n}$ that straddle the fractal set $S_c$ by $\phi_n$.  The number of refinements tested by the SGDA at step (iii) is constant $2^{n+k}= 2^{\bar n+\bar k}$, independent of $k$ or $n$. The SGDA integrates $\psi_n$ by a rectangle method using grid points, just as the BCA integrates $\phi_n$.  However, the SGDA uses {\it all} of the grid points at every step of the algorithm, only sorted differently, in order to get the $\psi_n$ estimates. The SGDA performs a total of $\bar k 2^{\bar n+\bar k}$ refinement tests, to be compared with the number of tests of the BCA which is only $2^{\bar n}(2^{\bar k-1}-1)$. Thus, the $\psi_n$ values are significantly more accurate than the $\phi_n$ values. We note that with increasing $n$ while keeping the same grid, the BCA performs better since values of $\phi_n$ with low $n$ and high error are removed from the least square fit toward the fractal dimension. Under the same conditions, the SGDA is expected to perform worse.  In this case, all $\psi_n$  values have comparable error, and eliminating some of them from the least square fit increases the error of the fitted slope. 

{\bf Numerics}. 
{\it (a) A piecewise linear dynamical system}. We consider the one-dimensional map \cite{TK} 
\begin{eqnarray}
\label{eq:ds}
g_{\mu,\nu}(x)=\left\{
\begin{array}{ll}
(x+1)/\mu-1, & x\leq -\nu (1-\mu)/(\nu+\mu);\\
-x/\nu, & |x|<\nu (1-\mu)/(\nu+\mu);\\
(x-1)/\mu+1, & \nu (1-\mu)/(\nu+\mu)\leq x;\\
\end{array}\right.
\end{eqnarray}
where $0<\nu <1$, and $0<\mu <1/2$. For $(\nu+2\mu)<1$, the dynamical system (\ref{eq:ds}) has two attractors 
at $\pm\infty$ that share a fractal BB. Here we choose $\nu=\mu$. The BB points are included in [-1,1], and the BB has a unique dimension which is $D=-\log 3/\log{\mu}$. 
Fig.~\ref{f:f2}(a) shows the box counting and the straddle grid dimension for $0.935<D<1$ (i.e., $0.31<\mu<0.3325$) where we used the Straddle test for refinements.  The black line represents $|D_g-D|$ versus $D$, and the gray line represents $|D_b-D|$ versus $D$, where $D_b$ is the box counting dimension.  We see that the SGDA performs significantly better than the BCA using the {\it same} grid of points.
\begin{figure}[t]
\begin{center}
\mbox{\epsfig{file=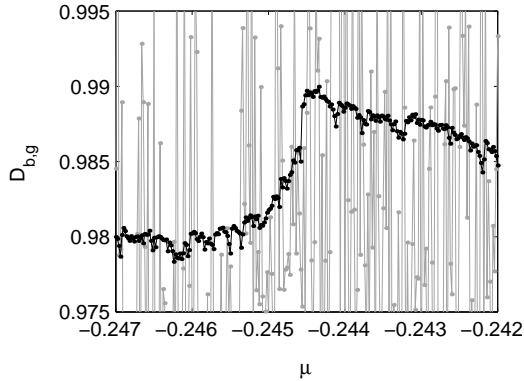,width=7cm}}
\caption{$D_{g}$ versus $\mu$ (black) and $D_{b}$  versus $\mu$ (gray), where $D_g$ ($D_b$) is the numericaly calculated grid (box) dimension, and $\mu$ is the forcing amplitude of the Duffing oscillator. We used $\bar n=1$, $\bar k=18$, and $r=1$. The refinements are tested with the BB straddle test. Again, our numerics clearly illustrate a considerable precision gain of SGDA as compared to BCA. Both algorithms use the same grid of points so they are equally fast. A jump in dimension is expected at $\mu\approx -0.2446$ due to a saddle-node bifurcation.  The graph of $D_g$  clearly illustrates this jump, while the graph of $D_b$ does not.}
\label{f:f3}\end{center}
\end{figure}
{\it (b) Nonlinear examples}. We consider the logistic map $f_\mu(x)=\mu x(1-x)$ for $0\leq x \leq 1$ and $\mu>4$; it is well-known that this system has a chaotic saddle. Figure \ref{f:f2}(b) shows $D_g$ (black) and $D_b$ (gray) versus $\mu$. The variation of $D_g$ has much less fluctuations than that of $D_b$, suggesting that the computation of $D_g$ is more accurate.

Our next example is constructed as follows.  Start with $f_\mu(x)$ for  $\mu=3.832$ where there is a stable period three orbit; denote its third iterate by $F(x)$. Consider now the map $h_\mu(x)=F(x)+\mu x e^{-7x}$. The map $h_\mu$ has fixed points attractors that share a fractal BB for $0<\mu<0.02$.  Figure \ref{f:f2}(c) shows $D_g$ (black) and $D_b$ (gray) versus $\mu$. The accurate dimension graph generated by the SGDA method provides interesting and useful information on certain bifurcations of $h_\mu$. For example, the jump in the dimension graph at $\mu\approx 0.01$ in Fig.~\ref{f:f2}(c) is caused by a saddle node bifurcation. 

The last example that we present is obtained as follows. Consider the forced Duffing oscillator \cite{Aguirre,Virgin,Breban} given by $d^2x/dt^2+0.15\, dx/dt +x-x^3=\mu\cos(t)$. Our dynamical system is the two-dimensional Poincar\'e map at constant phase of the forcing $t\mbox{ mod }2\pi=\pi$.  The map has attracting periodic orbits whose basins share a fractal boundary for $-0.247<\mu<-0.242$. In order to calculate the fractal dimension using SGDA, we choose a line segment $I$ in the phase space of the variables $x$ and $dx/dt$. All the points that belong to this segment have $dx/dt=1.35$ with $x$ ranging from $-2$ to $2$. The numerical results obtained for the fractal dimension by applying both BCA and SGDA are presented in Fig.~\ref{f:f3}.  Again, our numerics clearly illustrate a considerable precision gain of SGDA as compared to BCA. Both algorithms use the same grid of points so they are equally fast. A jump in dimension is expected at $\mu\approx -0.2446$ due to a saddle-node bifurcation which is similar to the one mentioned in our previous example.  The graph of $D_g$ (black) in Fig.~\ref{f:f3} clearly illustrates this jump, while the graph of $D_b$ (gray) does not.

{\bf Summary.} We introduced a novel method for computing the dimension of both fractal BB and the stable sets of chaotic saddles in supertransient systems. The SGDA is the first general, but simple method for computing the dimension of stable sets of chaotic saddles directly, fast and reliable. We expect the SGDA to be a useful additional tool in analyzing the dynamics of fluids, lasers and transition to turbulence in shear flows. 

{\bf Acknowledgments.} This work was done while H.E.N.~was visiting the Institute for Physical Science and Technology of the University of Maryland at College Park (Host: James A.~Yorke). It was in part supported by the NSF (Division of Mathematical Sciences and Physics, Grant No.~0104087).

\end{document}